\documentclass[sigconf, 10pt]{acmart}
\usepackage{amsmath,amsfonts}
\usepackage{algorithmic}
\usepackage{graphicx}
\usepackage{textcomp}
\usepackage{xcolor}
\usepackage{comment}
\usepackage{booktabs}
\usepackage[labelfont=bf]{caption}
\usepackage{subcaption}
\usepackage{enumitem}
\usepackage{tikz}
\usepackage{makecell}
\usetikzlibrary{decorations.pathreplacing,calligraphy}
\usepackage{multirow}
\usepackage{url}

\copyrightyear{2025}
\acmYear{2025}
\setcopyright{cc}
\setcctype{by}
\acmConference[SEC '25]{The Tenth ACM/IEEE Symposium on Edge Computing}{December 3--6, 2025}{Arlington, VA, USA}
\acmBooktitle{The Tenth ACM/IEEE Symposium on Edge Computing (SEC '25), December 3--6, 2025, Arlington, VA, USA}
\acmDOI{10.1145/3769102.3772714}
\acmISBN{979-8-4007-2238-7/2025/12}

\begin{document}

\title{A Distributed Framework for Privacy-Enhanced Vision Transformers on the Edge}

\author{Zihao Ding}
\affiliation{\institution{Rutgers University}
\city{Piscataway}
\state{NJ}
\country{USA}
}
\email{zihao.ding@rutgers.edu}

\author{Mufeng Zhu}
\affiliation{\institution{Rutgers University}
\city{Piscataway}
\state{NJ}
\country{USA}
}
\email{mz526@rutgers.edu}

\author{Zhongze Tang}
\affiliation{\institution{Rutgers University}
\city{Piscataway}
\state{NJ}
\country{USA}
}
\email{zhongze.tang@rutgers.edu}

\author{Sheng Wei}
\affiliation{\institution{Rutgers University}
\city{Piscataway}
\state{NJ}
\country{USA}
}
\email{sheng.wei@rutgers.edu}

\author{Yao Liu}
\affiliation{\institution{Rutgers University}
\city{Piscataway}
\state{NJ}
\country{USA}
}
\email{yao.liu@rutgers.edu}

\begin{abstract}
Nowadays, visual intelligence tools have become ubiquitous, offering all kinds of convenience and possibilities.
However, these tools have high computational requirements that exceed the capabilities of resource-constrained mobile and wearable devices. 
While offloading visual data to the cloud is a common solution, it introduces significant privacy vulnerabilities during transmission and server-side computation.
To address this, we propose a novel distributed, hierarchical offloading framework for Vision Transformers (ViTs) that addresses these privacy challenges by design.
Our approach uses a local trusted edge device, such as a mobile phone or an Nvidia Jetson, as the edge orchestrator.
This orchestrator partitions the user's visual data into smaller portions and distributes them across multiple independent cloud servers.
By design, no single external server possesses the complete image, preventing comprehensive data reconstruction.
The final data merging and aggregation computation occurs exclusively on the user's trusted edge device.
We apply our framework to the Segment Anything Model (SAM) as a practical case study, which demonstrates that our method substantially enhances content privacy over traditional cloud-based approaches. 
Evaluations show our framework maintains near-baseline segmentation performance while substantially reducing the risk of content reconstruction and user data exposure.
Our framework provides a scalable, privacy-preserving solution for vision tasks in the edge–cloud continuum.
\end{abstract}




\maketitle

\section{Introduction}
In recent years, visual intelligence tools have become increasingly ubiquitous, providing users with much convenience and capabilities through applications such as Google Lens, Microsoft Lens, and Apple's Visual Intelligence. These tools leverage advanced computer vision models to perform tasks such as image recognition, object detection, and image segmentation, enhancing the functionality and user experience of not only smartphones but also emerging wearable devices such as augmented reality (AR), virtual reality (VR), and mixed reality (MR) glasses.  
A defining feature of these devices is the always-on egocentric vision, which focuses on first-person perspective data from wearable devices, allowing the surroundings of a user to be captured and processed continuously.
For example, Meta's AR/MR glasses, including Ray-Ban glasses~\cite{metarayban}, Project Aria glasses~\cite{projectaria}, and their Orion prototype~\cite{projectorion}, are all equipped with cameras to support egocentric vision.
Similarly, Apple Vision Pro is a VR device with ``passthrough'' feature that can enable egocentric vision~\cite{apple2025immersive}. 
Egocentric vision has expanded the scope of visual intelligence applications, enabling more immersive and personalized user interactions.
As more and more smartphones and wearable devices come equipped with these out-of-the-box visual intelligence features, their adoption continues to surge, driven by the seamless integration of artificial intelligence into everyday applications.

While some visual intelligence tasks can be processed locally on the devices or offloaded to a nearby smartphone, more complex computations powered by massive foundation models, particularly Vision Transformers (ViTs)~\cite{dosovitskiy2020image}, often require offloading to more powerful machines equipped with high-end GPUs, due to the limited computation capability and battery power supply of mobile and wearable devices.
This may require sending the user's first-person view data to an external party, e.g., a cloud server, for processing.

The pervasive and always-on nature of egocentric vision, however, raises significant privacy concerns. 
For example, bystanders may feel uncomfortable about being recorded, whether intentionally or unintentionally, by the camera \\
wearer~\cite{plizzari2024outlook,denning2014situ,corbett2023bystandar}.
Egocentric images and videos can also contain sensitive personal information such as credit card numbers, addresses, computer screens displaying confidential data, or even images of the person's private space such as bedrooms and bathrooms~\cite{templeman2014placeavoider,korayem2016enhancing,roesner2014security,wu2025pipe}. 
Offloading egocentric vision tasks to cloud servers for processing can further increase users' privacy concerns, as private personal information can be vulnerable to breaches during data transmission and server-side computations.

The implications of compromised data privacy extend beyond mere information leakage. Attackers who gain access to the complete image data can exploit this information to create highly personalized and convincing scams using technologies like Deepfake~\cite{pei2024deepfake}. 
Numerous news reports highlight cases of people falling victim to Deepfake scams and and suffered significant financial losses~\cite{bbc-deepfake,cnn-deepfake}. 
U.S. Treasury Department recently released an alert highlighting the growing threat of such attacks~\cite{fincen-deepfake}. 
Research has also shown that images of the scene (e.g., the bedroom) can be reconstructed in high quality using only the extracted visual feature descriptors and/or point cloud 3D representations, even when the raw image data is deleted~\cite{pittaluga2019revealing,speciale2019privacy,dangwal2021analysis}.
These advanced attacks demonstrate an urgent need for better privacy protections in visual intelligence applications.  
To reduce these risks, it is essential to create new frameworks that offload and distribute computational tasks in a way that protects user data while maintaining the performance and responsiveness of visual intelligence applications.

% \vspace{0.3em}
In this paper, we propose a \textit{novel distributed hierarchical offloading framework} for enhancing content privacy in visual intelligence applications. 
This work is founded on our key insight that the window attention mechanism in modern Vision Transformers, originally introduced for computational efficiency, can be \textit{repurposed to enhance privacy in a distributed setting} ``for free.'' 
Our distributed framework is the first to exploit this insight, providing a practical, low-overhead solution that partitions data to limit exposure to any single untrusted party, with minimal impact on model utility and without incurring the high costs of traditional privacy techniques.

This partitioning is enabled by the hybrid architecture of state-of-the-art ViTs (e.g.,~\cite{kirillov2023segment,ravi2024sam,chen2024lw}). 
To handle high-resolution images, these models use computationally-efficient, window-based attention that processes image regions in isolation for the majority of transformer layers. 
Information integration across these separate windows is handled by a much smaller set of global attention layers.

Our system consists of three tiers: (1) a thin client (e.g., AR glasses) that captures data; (2) a trusted edge orchestrator (e.g., a smartphone with high-end mobile system-on-chip (SoC) or Nvidia Jetson); and (3) external cloud computing resources, 
Instead of sending the full visual data to a remote server for offloaded processing (e.g., image embedding extraction),
the edge orchestrator partitions user's visual data into smaller, non-overlapping pieces, and offloads the inherently parallel window attention computations to multiple, independent cloud servers. 
We assume these servers are honest-but-curious and non-colluding, a standard threat model detailed in Section~\ref{sec:threat_model}.
Results from these small data pieces are aggregated on the trusted local edge device. This aggregated data then passed through the global attention layers on the edge device  to produce the complete image embedding, which is then used for downstream vision tasks.
Since only limited private information (e.g., 1/25 of data in the image) is shared with each external resource, our design ensures that \textit{no single external computing party (i.e., the cloud server) possesses the complete image}.
Our partition-based scheme also prevents comprehensive data reconstruction even in the event of a data breach or unauthorized access, enhancing the privacy of user data. 
Meanwhile, enhancing privacy should not sacrifice the utility of vision tasks. 
Our comprehensive evaluation results show that image embeddings extracted from our privacy-enhanced ViT framework can achieve similar vision task performance as conventional approaches.
In summary, this paper makes the following contributions:

\begin{itemize}[leftmargin=*,topsep=3pt]
\item We design a privacy-enhancing hierarchical offloading framework for ViT-based visual intelligence models, leveraging distributed computation on local trusted edge devices and multiple external parties.
\item We propose a practical, application-specific, methodology for evaluating visual privacy risks against state-of-the-art reconstruction and detection adversaries.  
\item We implement and deploy
our privacy-enhanced distributed computation
framework on a state-of-the-art image segmentation model, the
Segment Anything Model (SAM)~\cite{kirillov2023segment}, to demonstrate its feasibility using Docker and gRPC for edge-cloud communication.
\item Extensive evaluation results show our framework largely maintains high task accuracy while effectively protects user privacy.
Even with state-of-the-art strong adversaries, limited pixel-level and object-level information can be recovered from the small, sub-portion of the image that is processed at external computing resources.
\end{itemize}

\section{Background and Related Work}

\subsection{Visual Privacy Protection}
Initial approaches to visual privacy have relied on simple obfuscation techniques such as blurring and pixelation.  
For example, EgoBlur~\cite{raina2023egoblur} protects bystanders' faces and vehicle license plates by using FasterRCNN~\cite{ren2016faster} for object detection and obscuring the corresponding regions. 
BystandAR~\cite{corbett2023bystandar} labels faces in the frame as a ``subject'' or a ``bystander'' of the conversation based on the camera wearer's gaze and voice information and obscures bystanders' faces. 
These two solutions, however, focus on bystander privacy and do not protect other sensitive personal identification information. 
The computation required for these solutions can also be prohibitively high for mobile devices~\cite{raina2023egoblur,corbett2023bystandar}.
Furthermore, this type of obfuscation removes fine detail needed for precise vision tasks and is increasingly vulnerable to reversal by modern restoration/inpainting or recognition models~\cite{mcpherson2016defeating}.

Protective perturbation~\cite{ye2022visual} obfuscates sensitive data by adding perturbations with the goal of making images unrecognizable to humans while preserving the accuracy of desired machine vision tasks.
Separately, NinjaDesc~\cite{ng2022ninjadesc} uses an adversarial learning framework to train a descriptor encoder network that generates content-concealing visual descriptors intended to prevent image reconstruction. 
Both works use adversarial learning for proactive privacy protection. 
However, protective perturbation requires training a dedicated perturbation generator for each target model it aims to support. 
NinjaDesc requires retraining its encoding network for each type of base descriptor it needs to conceal. These requirements limit the portability and increase deployment cost.
Moreover, for these approaches, a powerful enough model trained on sufficient pairs of  original and perturbed data could potentially learn to reverse the transformation. 
Such models have seen success in image generation. For example, diffusion models~\cite{ho2020denoising,zhu2023denoising} and vision transformers~\cite{yu2022scaling} allow decoding image content from information-rich latent representations. 
Although not directly aligned with this obfuscated representation decoding problem, as long as the transformed representation does not discard significant information, then given enough examples, we would expect standard decoding models to reconstruct original imagery from a perturbation-encoded representation. 

Beyond vision-specific obfuscation and descriptor learning, a number of system-level privacy mechanisms have been proposed for inference on remote servers.
Homomorphic Encryption (HE) and Secure Multi-Party Computation (MPC) offer formal cryptographic guarantees, but they cannot be directly applied to unmodified Vision Transformers: nonlinear activations and attention layers must be approximated or replaced by protocol-friendly primitives, and current HE/MPC pipelines report substantial latency and communication overhead in realistic network settings (e.g., CrypTFlow2~\cite{rathee2020cryptflow2} and Cheetah~\cite{huang2022cheetah}).
Recent co-design efforts (e.g., activation-function co-design for MPC~\cite{diaa2024fast}) can reduce those overhead but at the cost of substantial modifications to both model architectures and cryptographic protocols. 
Differential Privacy (DP) can be applied with fewer architectural modifications by injecting calibrated noise at inputs or intermediate features.
However, downstream task performance degrades as noise increases unless the model is retrained for robustness (e.g., PixelDP-style test-time noise~\cite{lecuyer2019certified}). 
In contrast, our distributed ViT framework avoids both the heavy cryptographic overhead and the direct utility loss from input perturbation by preserving full pixel fidelity within each partition while limiting any single server's view under a semi-honest, non-colluding model; the ViT runs unchanged, and the primary costs are modest edge-side coordination and communication for partitioned processing.

\begin{figure*}[!h]
\centering
\includegraphics[width=0.95\textwidth]{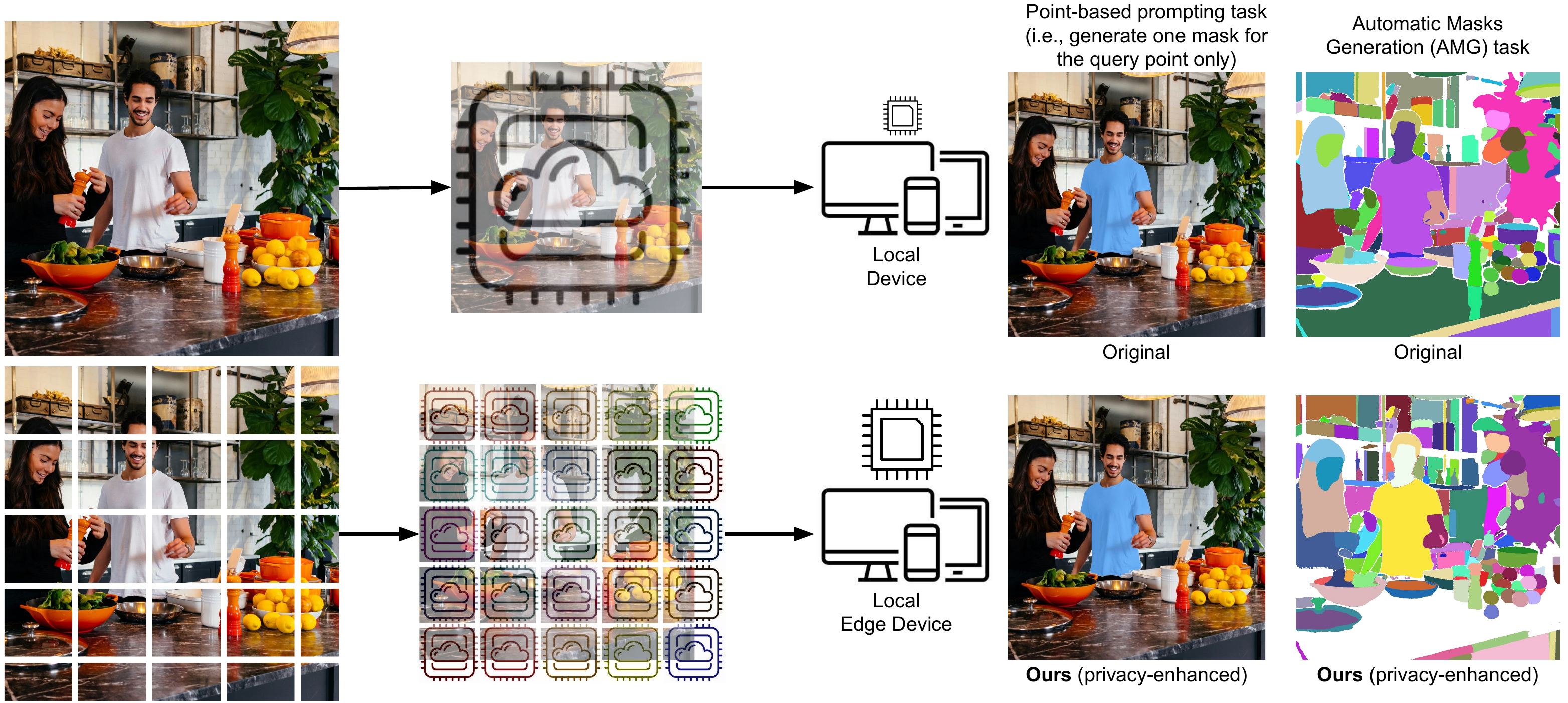}
\vspace{-0.2em}
\caption{\textbf{Top:} \textmd{Typical cloud-based visual intelligence task workflow: offload the entire image to a cloud server for generating the image embedding. 
The image embedding is then used by the local device via lightweight computation for downstream vision tasks.}
\textbf{Bottom:} \textmd{Our proposed privacy-enhanced solution: partition the entire image into $w\times w=w^2$ windows (e.g., 25 shown in the figure). Content in each window is processed separately by a different external privacy domain (e.g., a cloud service provider) to extract per-window embeddings. These window embeddings are then merged and further processed at a local trusted edge device to obtain the final image embedding. 
The extracted image embeddings are combined with a prompt encoder and a mask decoder for image segmentation tasks. \textit{The four figures on the right show example output of 2 different use cases: (i) mask generation for a query point; and (ii) automatic masks generation (a difficult task). The results show that the image embedding generated by our privacy-enhanced solution can create the same single mask generation output as the original approach; and can perform the more difficult masks generation task well with only slight performance decrease.}}}
\vspace{-0.5em}
\label{fig:poc-results}
\end{figure*}

\subsection{Typical Cloud-Based Visual Intelligence Task Workflow} 
Figure \ref{fig:poc-results} (Top) shows an example cloud-based visual intelligence task workflow. The local device offloads the compute-intensive visual task by uploading the visual data (e.g., an image) to the cloud server for heavyweight processing, which returns an image embedding to the user's local device. With the image embedding, only lightweight computation is needed for performing the desired vision tasks. 

Two tasks are shown in the figure: an image segmentation task with ``point-based prompting'' that takes the coordinates of a point as input and returns a mask for all pixels belong to the same object as the point prompt, e.g., the white shirt is masked in blue color in the figure; and an ``automatic masks generation'' image segmentation task that generates all reliable masks in an input image, e.g., each object is masked with a different color in the output image. 
However, with this workflow, 
\textit{the full image is uploaded to a cloud service server for processing, posing privacy risks}.
Even if the full image can be perturbed or blurred, it can be reconstructed via state-of-the-art image recovery~\cite{orekondy2017towards,zhu2023denoising}.

\subsection{Vision Transformer (ViT)} 
Since its introduction, Vision Transformer (ViT)~\cite{dosovitskiy2020image} and its variants (e.g., Swin Transformer~\cite{liu2021swin,liu2022swin}) have been shown to be successful on various vision tasks, such as image classification~\cite{dosovitskiy2020image,he2022masked}, object detection~\cite{li2022exploring}, image restoration~\cite{liang2021swinir,conde2022swin2sr},  image segmentation~\cite{kirillov2023segment,ravi2024sam}, hand and object tracking~\cite{banerjee2025hot3d}, pose estimation~\cite{xu2022vitpose}, etc. 

\begin{figure}[!t]
\centering
\begin{subfigure}[t]{0.23\textwidth}
\begin{tikzpicture}[thick]
  \node[anchor=south west, inner sep=0] (image) at (0,0) {\includegraphics[width=\textwidth]{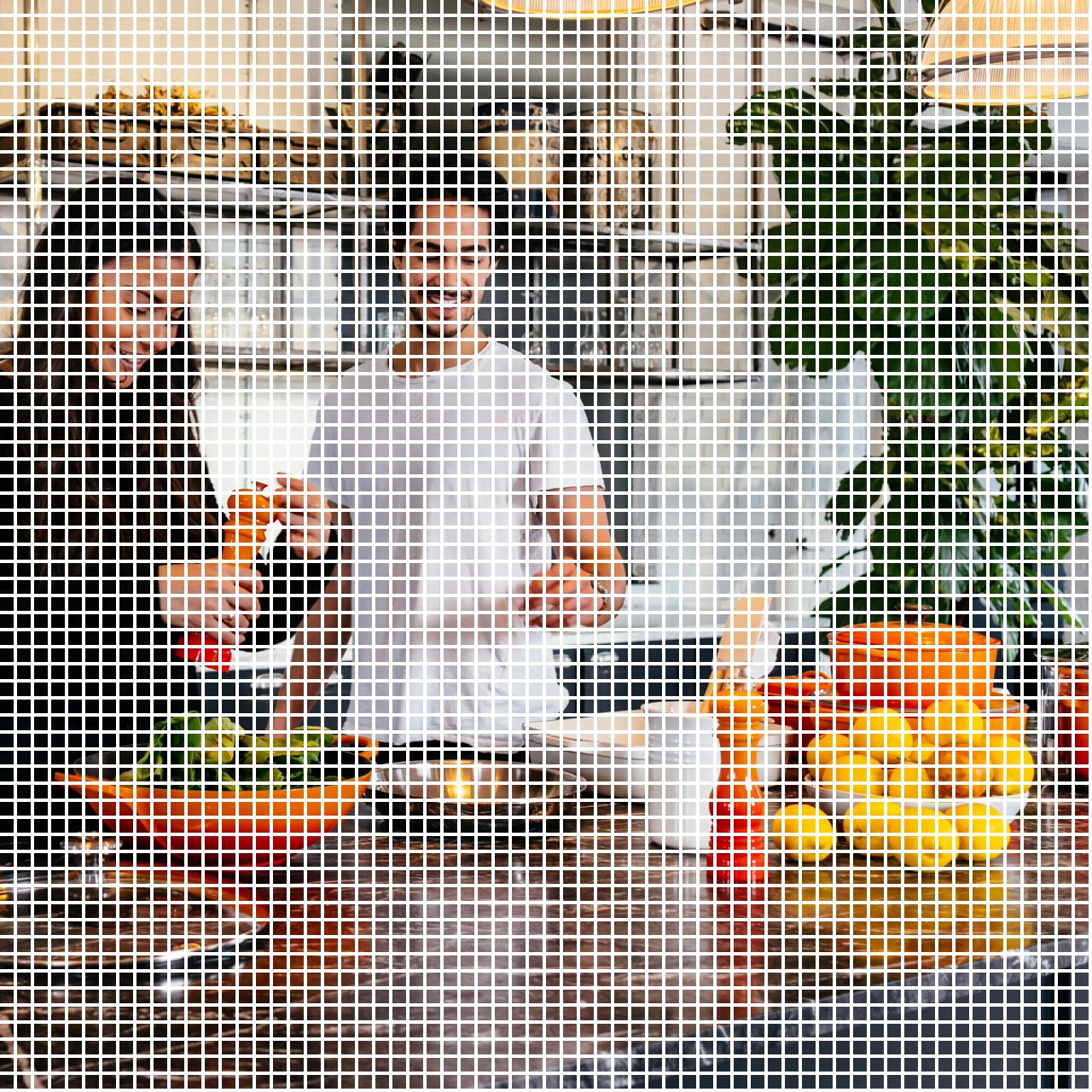}};
    \draw [decorate,decoration={brace,amplitude=6pt,mirror,raise=7pt},yshift=2.2cm] (4.1,1.8) -- (0,1.8) node [black,midway,yshift=16pt] {$n$};
\end{tikzpicture}
\caption{Global attention in ViT attends to all $n^2$ patches in the full image.}
\end{subfigure}
\hfill
\begin{subfigure}[t]{0.23\textwidth}
\begin{tikzpicture}[thick]
  \node[anchor=south west, inner sep=0] (image) at (0,0) {\includegraphics[width=\textwidth]{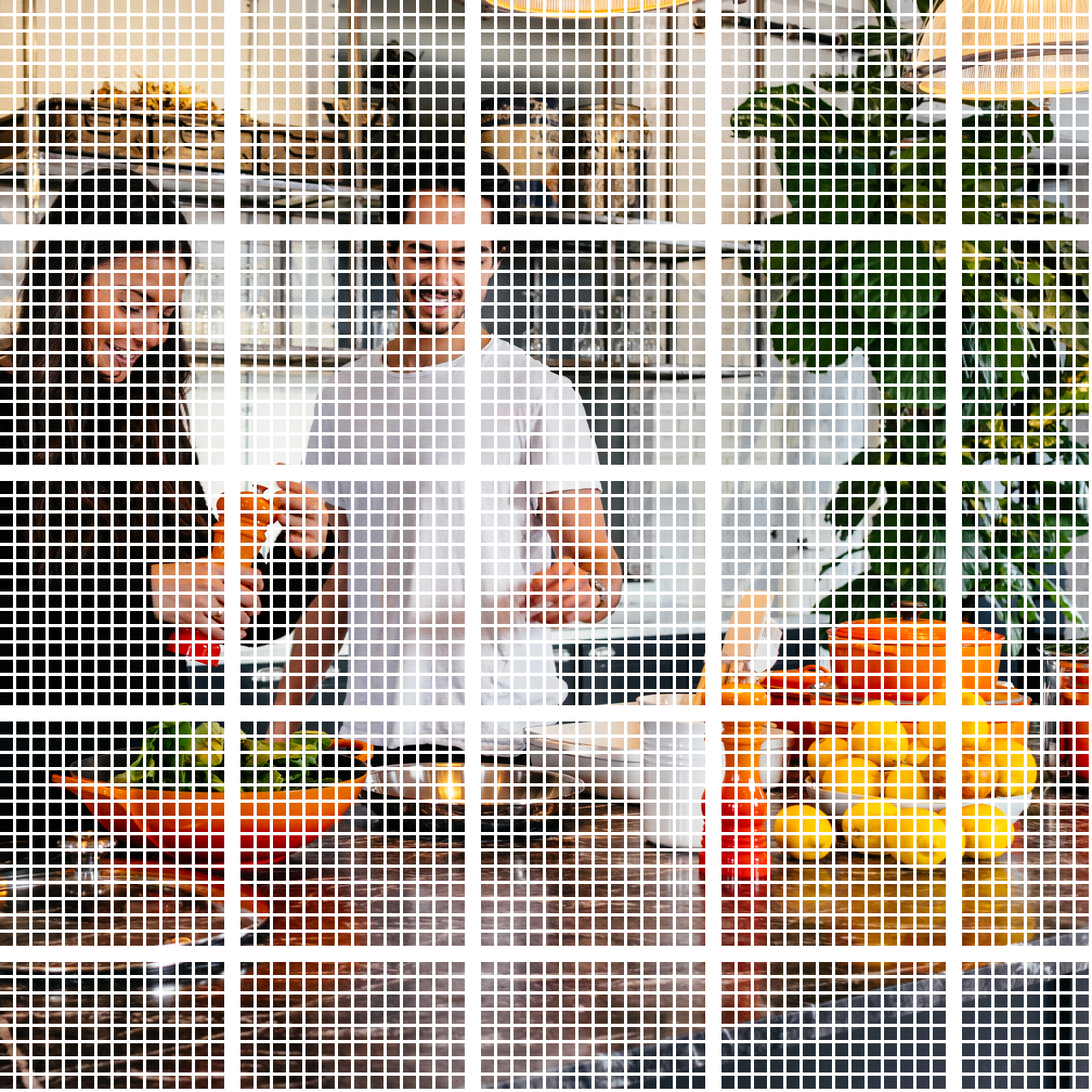}};
    \draw [decorate,decoration={brace,amplitude=5pt,mirror,raise=7pt},yshift=2.2cm] (0.85,1.8) -- (0,1.8) node [black,midway,yshift=16pt] {$r$};
\end{tikzpicture}
\caption{Window attention in ViT attends to all $r^2$ patches in a window.}
\end{subfigure}
\vspace{-0.8em}
\caption{
\textmd{
Existing ViT-like models use a combination of global attention and window attention blocks for extracting an image embedding for downstream vision tasks.
}}
\vspace{-1.5em}
\label{fig:global_local}
\end{figure}

With ViT-like models, an input image is first rescaled to a resolution supported by the Transformer model and
then divided into patches, each with $16\times 16$ pixels. Overall, $n\times n$ patches are created, as shown in Figure \ref{fig:global_local}(a). 
Each $16\times 16$ pixels patch is linearly projected to a patch embedding of dimension $d$, e.g., via a 2D convolution kernel with the same size as the patch itself. 
A typical self-attention layer (also referred to as \textbf{ global attention}) in ViT attends to the full flattened sequence length of $n^2$, i.e., the total number of patches in the image. 
Figure \ref{fig:global_local}(b) shows an example of \textbf{window attention}, where the full set of patches are partitioned into windows, with each window containing $r\times r$ patches. 
Self-attention is only computed within the $r\times r$ patches in the window.
Window attention is often used for reducing computational costs of attention calculations.
Information can be propagated across windows via the global attention layers only. 
An example ViT-like model can include a number of self-attention blocks. 
For example, the image encoder of the Segment Anything Model (SAM)~\cite{kirillov2023segment} is based on ViT-Huge (ViT-H). It includes 32 self-attention layers that are divided into 4 groups. 
Each group include 7 window attention layers followed by 1 global attention layer.
The 28 window attention layers account for  84.8\% floating-point operations (FLOPs) of the total FLOPs across the 32 layers, with the remaining computation  attributed to the 4 global attention layers.

\section{A Practical Metric for Visual Privacy}

We first propose a practical metric for computing visual privacy risks. 
The metric is application-specific.
For a target application, we define a prediction problem that takes external information (e.g., the training dataset of the prediction model, or a database of people's pictures or vehicle registration information) and a portion of the image (i.e., information made available to one external privacy domain). 
It tries to predict private information (e.g., identity of a person or a vehicle in the image or pixels in the original image).
Our proposed infrastructure will use state-of-the-art prediction approaches specific to the application. 

Given an original image $\mathfrak{I}$, we rely on cloud server $s$ for processing a small sub-portion of the image, $\mathfrak{i}^{(s)}$. An adversary at the cloud server can use a state-of-the-art prediction approach, denoted as $\mathbf{F}(\cdot)$, for deriving private information from the portion of the image  $\mathfrak{i}^{(s)}$, obtaining $\mathbf{F}(\mathfrak{i}^{(s)})$,  a distribution over sensitive information.

To assess privacy risks, we are also assuming an application-specific function $\mathbf{PI}(\cdot)$ that can assess a true set of private information within the full input image $\mathfrak{I}$.
For example, 
for an application requiring pixel-level privacy, private information includes the image itself;
for object-level privacy, private information includes semantic information such as the types of objects in the image and their positions.
We can then \textit{compute visual privacy risk} $VPR$ as a function of the output given by the adversary's prediction $\mathbf{F}(\mathfrak{i}^{(s)})$ and the groundtruth private information in the image $\mathbf{PI}(\mathfrak{I})$.
That is, $VPR(\mathfrak{I}, \mathfrak{i}^{(s)}) \approx VPR^{\texttt{(app-specific)}} (\mathbf{F}(\mathfrak{i}^{(s)}), \mathbf{PI}(\mathfrak{I}))$.
This privacy risk measurement is intended to be aligned with the expected amount of sensitive information that an adversary can predict from portions of an image sent to external privacy domains.
We acknowledge that application-specific privacy risk may change with advances in current prediction approaches. 

\vspace{0.3em}
\noindent\textbf{Example pixel-level privacy.}
We assume that for our application, 
sensitive information consists of pixel-level image data. 
In this case, a strong adversary could use a pixel-level prediction approach to recover sensitive information. 
For example, the adversary could use a state-of-the-art masked autoencoder (ViTMAE)~\cite{he2022masked} to infer missing part of the image. ViTMAE was designed to train a strong ViT encoder backbone. 
Inspired by BERT~\cite{devlin2018bert}, it uses a self-supervisory task that masks the majority of the image and train a heavyweight encoder that can predict the missing parts of the image via a lightweight decoder.
ViTMAE has been shown to predict plausible information in an image, even under more than 75\% mask ratio.
The prediction performance of ViTMAE differs under different masking strategies, with the ``random'' and ``grid'' strategies resulting in the best prediction results (good visual results under 75\% mask ratio) and the ``block'' being the worst (bad visual results with just 50\% mask ratio) (Figure 6 in \cite{he2022masked}). In other words, if ViTMAE is just given a patch of the image, it is very hard to correctly predict other portions of the full image.
We can then evaluate application-specific privacy risk by assessing the pixel-level loss of these ViTMAE methods, e.g., PSNR, structural similarity (SSIM)~\cite{wang2004image}, and perceptual similarity metric, LPIPS~\cite{zhang2018unreasonable}.

\vspace{0.3em}
\noindent\textbf{Example object-level privacy.}
For object-level privacy, we assume that sensitive information consists of semantic information contained in the images, e.g., the types of objects and the size and location their bounding boxes in the image. 
In this case, an adversary can run state-of-the-art
deep neural network models such as YOLOv9~\cite{wang2024yolov9}, on
the partial information provided in $\mathfrak{i}^{(s)}$ 
or on the image restored by MAE, to obtain object detection results. 
We can then evaluate the visual privacy risk by comparing these results with the results obtained from the original image $\mathfrak{I}$.

\section{Threat Model}
\label{sec:threat_model}
In this work,
we model the external cloud servers as \emph{honest-but-curious} and \emph{non-colluding adversaries}. 
This standard assumption dictates that each server correctly executes the protocol but may attempt to infer sensitive information from the data it processes (its assigned windows), without conspiring with other servers.
A trusted edge orchestrator performs partitioning and computes global attention locally. Raw images and global-attention intermediates never leave the edge, and edge–cloud links use secure channels. The adversary’s goals are (i) \emph{pixel-level reconstruction} of hidden content from a single window and (ii) \emph{object-level inference} within a window. 
We therefore measure reconstruction quality and post-reconstruction detection accuracy in our evaluation. 
Similar semi-honest, non-colluding models are common in hierarchical/edge private inference and two-server MPC (e.g., ~\cite{dehkordi2024privatemdi,gu2021precad}), and are standard in MPC foundations ~\cite{evans2018pragmaticmpc}.

\begin{figure*}[!t]
    \centering
    \includegraphics[width=0.85\textwidth]{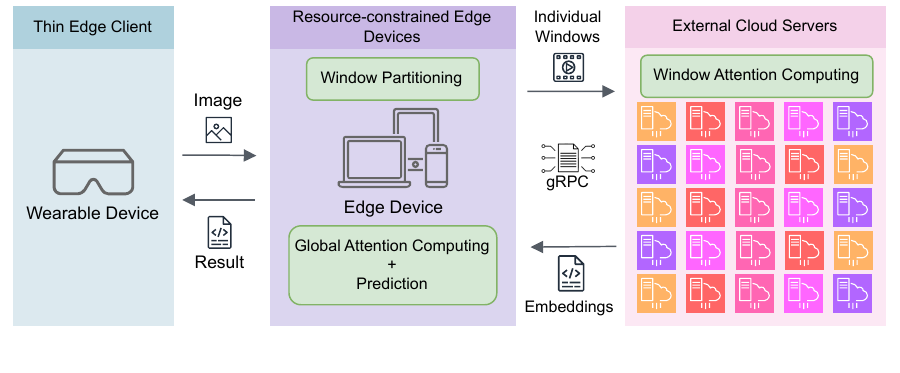}
    \vspace{-2.9em}
    \caption{\textmd{Our proposed hierarchical offloading framework enhances privacy in visual intelligence applications by using a local trusted edge orchestrator (e.g., an Nvidia Jetson) to partition data from a thin edge client and offload the computation across external cloud servers operated by different administrative domains. This design ensures that no single external party processes the complete data.}}
    \label{fig:system-architecture}
\vspace{-0.5em}
\end{figure*}

\section{Design}

We design a distributed, hierarchical offloading framework for enhancing the visual privacy of user data in ViT-based applications.
Figure \ref{fig:system-architecture} shows the workflow of our proposed privacy-enhanced distributed computation framework.  
At a high-level, the edge device acts as an orchestrator, which partitions the user's data (e.g., an image) into multiple, non-overlapping sub-portions. Each portion is then offloaded to an external computing resource (e.g., different cloud service providers) for processing. The resulting embeddings are sent back to the edge orchestrator for the final global attention computation and prediction to generate the final result. 
We assume that these external computing resources are owned and managed by different administrative domains (e.g., different companies, organizations, or even different users) and do not collude with each other.
This partitioned approach ensures no single external party gains access to the complete data, thus enhancing the visual privacy. 

\subsection{Distributed Computation Framework for ViT}

To support the partition-based, privacy-enhanced workflow for ViT processing, 
existing vision model components/layers must be adapted to facilitate processing a small ``window'' of the image. To do so, 
our method leverages the inherent characteristics of Transformer's  window (local) and global attention mechanisms~\cite{vaswani2023attentionneed} to distribute computational tasks from local devices to cloud servers. 
We make use of the \textbf{window attention} (recall Figure \ref{fig:global_local}(b)) mechanism for processing data within each window only on an external computation resource (e.g., a cloud resource). 
The outputs of all windows' window attention layers are returned back to the local device, which are \textbf{merged} and processed through the global attention layers on a local device to obtain the image embedding of the full image. 
Overall, our distributed computation framework include the following key components:

\vspace{1mm}
\noindent\textbf{Window-based image partitioning.}
The input image is divided into non-overlapping windows. Each window serves as an independent \textit{unit of computation}. This partition ensures that no single partition contains sufficient information to reconstruct the entire image, therefore enhancing privacy.

\vspace{1mm}
\noindent\textbf{Remote window-based processing.}
Instead of processing all windows locally, each window is sent to a remote server for computation. With parallel processing of image partitions across different servers, each server processes the received windows through preloaded ViT blocks and returns the processed results to the requesting client (i.e., the local device). 
By distributing the processing load, the framework achieves parallelism, enhancing computational efficiency, and reducing the overall processing time. This design also ensures scalability, as additional servers can be incorporated to handle increased workloads or higher-resolution inputs, enabling the system to adapt seamlessly to varying demands.

\vspace{1mm}
\noindent\textbf{Local full image processing.}
The local trusted edge device collects all extracted window-based embeddings and merges them. 
The merged representation is further processed by the final global attention layers at the edge device, 
integrating information across windows in the entire image to obtain the overall image embedding. 
This step ensures that while window-based computations are distributed, the global context necessary for further computation remains local.

A natural concern with such a partition-based approach is if it can achieve good vision task performances. After all, cross-window information cannot be extracted during the window-based processing stage at the cloud servers and can only be obtained via the global attention layers in the full image stage. 
Here, we cite an important observation made in ViTDet~\cite{li2022exploring}. 
ViTDet is an object detector that uses plain ViT backbones. 
For reducing the computation and memory requirement of the attention layers, ViTDet mainly uses window attention layers with a few cross-window global attention layers. 
While evenly placing the global attention layers can achieve the best results, they found that placing all the global attention layers in the last few layers (e.g., the last 4 layers in their paper) can achieve results that are almost as good (Table 2(c) in \cite{he2022masked}).
Inspired by this observation, we apply our partition-based, privacy-enhanced distributed computation framework to a state-of-the-art image segmentation model, the Segment Anything Model (SAM) by Meta~\cite{kirillov2023segment}.

\begin{figure*}[!t]
\centering
\begin{subfigure}[b]{0.46\textwidth}
    \centering
    \includegraphics[width=0.93\textwidth]{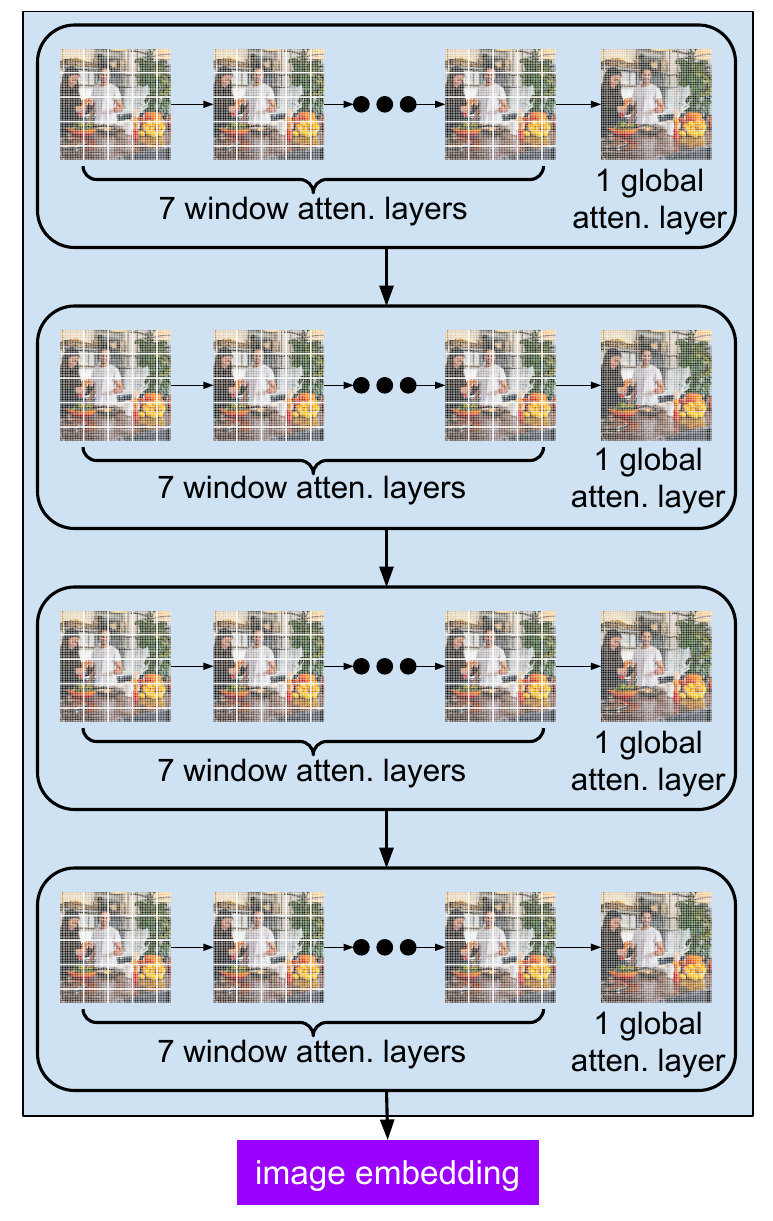}
    \caption{Image encoder of the original Segment Anything Model (SAM-H) with 32 atten. layers. }
\end{subfigure}
\hfill
\begin{subfigure}[b]{0.49\textwidth}
    \centering
    \includegraphics[width=\textwidth]{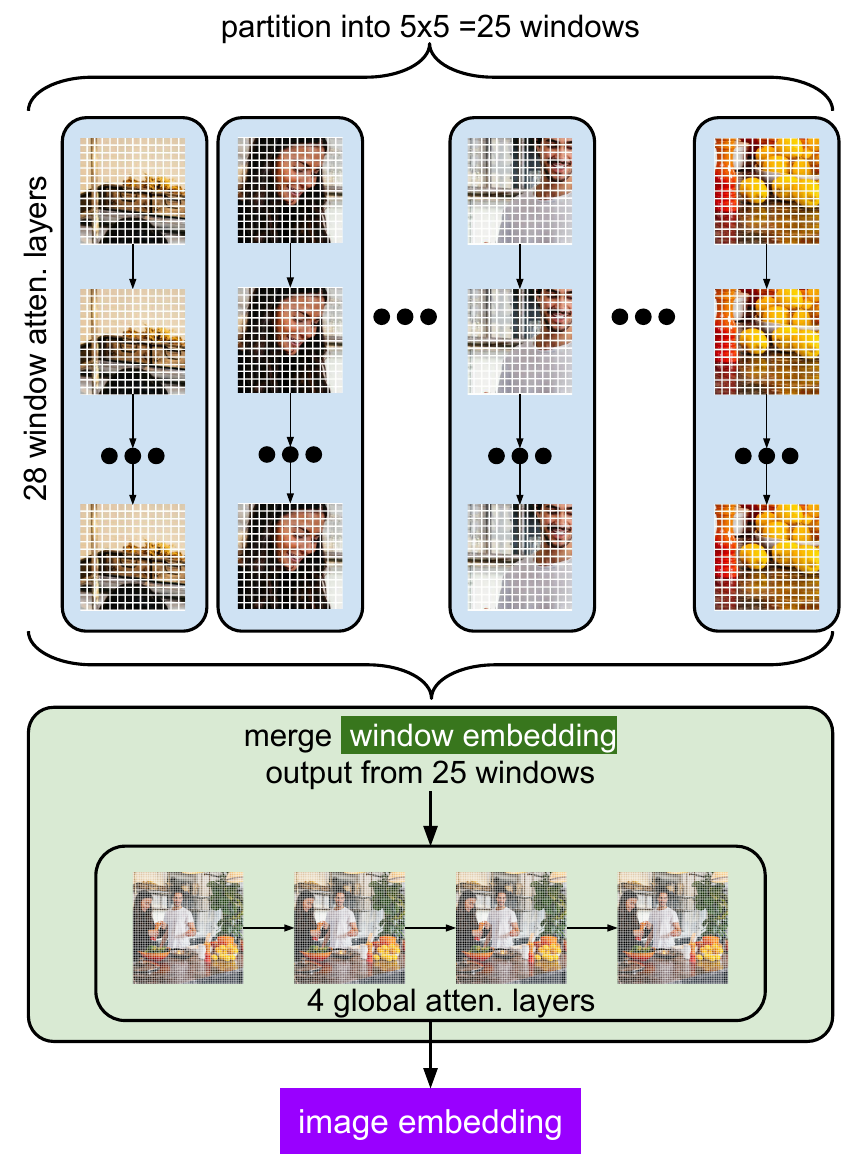}
    \caption{Our proposed privacy-enhanced solution with separated window atten. and global atten. layers. }
\end{subfigure}
\vspace{-0.5em}
\caption{\textbf{(a)} \textmd{In the original Segment Anything Model (SAM-H), 32 layers of attention are divided into 4 groups. Each group include 7 window attention layers followed by 1 global attention layer. Typically, the full image encoder (in blue-shaded box) executes at the cloud server for extracting the image embedding from an input image. }
\textbf{(b)} \textmd{In our privacy-enhanced solution, an image is partitioned into 25 windows. Content of each window is processed via a third-party, external, cloud provider (thus 25 parties overall) for extracting features local to the window via 28 window attention layers, shown in blue-shaded box. Their outputs are then transmitted, received, and merged  at the edge device and further processed by 4 global attention layers (shown in green-shaded box). 
In both figures, the multi-layer perceptron (MLP) layer  operations at the end of each attention layer and 2D convolution operations at the end of the image encoder are omitted.}}
\label{fig:poc}
\end{figure*}

\subsection{Example Vision Task: Segment Anything Model (SAM)}
The original SAM model includes three main components: a heavyweight \textit{image encoder}, a lightweight prompt encoder, and a lightweight mask decoder. 
The \textit{image encoder} uses a ViT-like architecture to create
\textit{image embeddings}
and only needs to run once. It adapts the ViT model to support higher-resolution images (e.g., $1024 \times 1024$ in SAM models vs. $224\times 224$ in ViT models.)
The image embedding combined with prompts encoded by the prompt encoder are processed by the mask decoder, which generates the desired segmentation mask. 
Figure \ref{fig:poc}(a) shows the architecture of the ViT-based image encoder used in SAM. 
Here, in the SAM model with ViT-Huge (SAM-H), the ViT-H image encoder includes 28 window attention layers and 4 global attention layers, and the global attention layers are evenly placed among all layers in the ViT.
Besides ViT-H, the SAM model also supports two image encoders with smaller ViT backbone sizes: ViT-Large (ViT-L) and ViT-Base (ViT-B).

We adapt all three SAM-ViT backbone models to our proposed partition-based, privacy-enhanced model. We refer to them as Privacy-Enhanced Distributed Segment Anything Model (PED-SAM).
Figure \ref{fig:poc}(b) shows the adapted model for the original SAM ViT-H model.
Given the 32 attention layers, we use the first 28 layers as window attention layers and the last 4 layers as global attention layers.
Note that regardless if it is a window attention layer or a global attention layer, the number of model parameters in each layer is the same. 
Furthermore, \textit{we do not retrain or finetune the model}. Instead, we simply use original model weights to initialize the weights in each layer, even if the corresponding layer has been changed from a global layer to a local layer or vice versa. 

Since the first 28 attention layers process content within a window only, we can partition the input image into $w^2$ windows (25 as shown in Figure \ref{fig:poc-results} (Bottom)), each processed by a different cloud resource. 
Overall, computing the window embeddings for all windows accounts for more than 84\% of the floating-point operations (FLOPs) for computing the image embedding. 
That is, more than 84\% of the computations are offloaded to cloud servers in our privacy-enhanced distributed computation framework.
The window embedding outputs are aggregated and further processed by the global attention layers to obtain image embeddings at the local trusted edge device. 
We then use the extracted image embeddings for segmentation tasks. 
Even with such naive model adaptation, we show in Section \ref{sec:eval} that our privacy-enhanced model can still achieve results that are nearly as good. More visual results can be found in Figure \ref{fig:poc-results}.

\section{Implementation}

Our implementation is based on the PyTorch implementation of the SAM model. The workflow began with preprocessing the input images by dividing them into smaller, non-overlapping windows. Each window embedding was treated as an independent computational unit. These serialized windows were transmitted to remote servers via gRPC, where window attention computations were performed using preloaded Transformer blocks within the SAM architecture. Once the servers processed the windows, the computed window-embeddings are returned via gRPC to the edge device. The edge device reassembles them by performing window merging, followed by processing them through the final global attention layers.
This design offloads computationally intensive tasks (e.g., more than 84\% FLOPs) to remote servers while ensuring that the complete image context remains local.

To setup runtime environments for local edge devices and remote servers, we used Docker containers. 
To model different types of edge hardware, we configured two types of containers for edge devices: \textbf{CPU-only} containers to emulate resource-constrained edge devices, and \textbf{GPU-enabled} containers to represent more powerful edge devices.
In contrast, the cloud server containers have accesses to high-end GPUs, representing the higher computational power of cloud servers. This configuration facilitated easy deployment across different devices and accurately reflected the resource constraints typically encountered in edge-cloud computing scenarios.

The communication between the edge device and servers was facilitated through \textbf{gRPC}~\cite{grpc}, a high-performance, open-source framework enabling efficient and low-latency data transmission. gRPC enabled the parallel processing of image partitions across multiple servers.

\section{Evaluation}
\label{sec:eval}
In this section, we evaluate the performance of our Privacy-Enhanced Distributed Segment Anything Model (PED-SAM), an adaptation of the Segment Anything Model (SAM) integrated with our proposed privacy-enhanced distributed ViT framework. The evaluation focuses on two key aspects: utility and privacy. 
Through systematic experiments, we examine the model's computational efficiency, vision task accuracy, and its ability to safeguard user content against image reconstruction attacks in the event of data leaks or breaches.

\subsection{Utility Evaluation}
\noindent\textbf{Experimental Methodology.}
This evaluation quantifies the performance of our Privacy-Enhanced SAM (PED-SAM) framework. Using the ViT-B SAM checkpoint on the COCO val2017 benchmark~\cite{lin2014microsoft}, we systematically measure end-to-end latency to demonstrate how our two-stage architecture leverages distributed compute to accelerate inference while preserving segmentation utility.

Our distributed architecture is composed of two primary roles: the \emph{Resource-constrained Edge
Devices}, which runs the main application, executes the final global-attention and predictor stages, and orchestrates offloading; and the \emph{External Servers}, which receive and process the computationally intensive window-attention tasks in parallel.
To test our framework across a spectrum of real-world hardware, our testbed instantiates these roles with a variety of platforms. The \textit{Resource-constrained Edge
Devices} role is fulfilled by:
\begin{itemize}
\item \textbf{A High-End Desktop:} An Alienware R14 (AMD Ryzen 9 5900X, NVIDIA RTX 3080 Ti).
\item \textbf{A High-Performance Edge SoC:} An NVIDIA Jetson AGX Orin 64GB module.
\item \textbf{A Resource-Constrained Edge SoC:} An NVIDIA Jetson Orin Nano 8GB module.
\end{itemize}
The \textit{External Server}'s role is fulfilled by a heterogeneous pool of three GPUs: two NVIDIA RTX 4090s, each given dedicated PCIe-passthrough access within the Proxmox VM, and one NVIDIA RTX 3080 Ti hosted in a separate Dockerized workstation.

%=========== TABLE 1: CPU-BASED ================
\begin{table*}[!t]
\centering
\caption{Performance comparison across different hardware platforms when edge devices are configured with \textbf{CPU-only containers}. We report the average latency for the image encoder and the mask predictor. All times are measured in milliseconds (ms). Our privacy-enhanced methods show significant latency reduction on edge devices through distributed computing, with minimal overhead in local configurations. 
\textbf{Abbreviations}: Enc. (Encoder), Pred. (Predictor), J5 (JetPack 5), J6 (JetPack 6).}
\label{tab:cpu_perf_comparison}
\setlength{\tabcolsep}{4.5pt} 
\renewcommand{\arraystretch}{1.15} 
\begin{tabular}{@{}l
            r@{\hspace{0.8em}}r
            r@{\hspace{0.8em}}r
            r@{\hspace{0.8em}}r
            r@{\hspace{0.8em}}r@{}}
\toprule
\multirow{2}{*}{\makecell[l]{\textbf{SAM Configuration} \\ \textbf{(CPU-based)}}} &
\multicolumn{2}{c}{\textbf{{\makecell{Jetson Orin\\ Nano 8GB}}}} &
\multicolumn{2}{c}{\textbf{{\makecell{Jetson AGX Orin\\(J5, 1-core)}}}} &
\multicolumn{2}{c}{\textbf{{\makecell{Jetson AGX Orin\\(J6, 12-core)}}}} &
\multicolumn{2}{c}{\textbf{Ryzen 9 5900X}} \\
\cmidrule(lr){2-3}\cmidrule(lr){4-5}\cmidrule(lr){6-7}\cmidrule(lr){8-9}
& \textbf{Enc.} & \textbf{Pred.} & \textbf{Enc.} & \textbf{Pred.} & \textbf{Enc.} & \textbf{Pred.} & \textbf{Enc.} & \textbf{Pred.} \\
\midrule
Original SAM (Baseline)              & 17,698 & 444 & 419,242 & 2,312 & 7,899 & 188 & 4,135 & 46 \\
\midrule
\textit{Privacy-Enhanced SAM (Local Execution)}    & 17,700 & 426 & 443,747 & 2,330 & 8,239 & 254 & 4,160 & 48 \\
\midrule
\multicolumn{9}{@{}l}{\textit{Privacy-Enhanced SAM (Distributed Execution with Remote GPUs)\textsuperscript{a}}} \\
\quad Offload to 1x GPU (1x RTX 4090)           & 9,923 & 369 & 210,345 & 2,335 & 4,105 & 257 & 3,434 & 48 \\
\quad Offload to 2x GPUs (2x RTX 4090)          & 9,715 & 391 & 209,024 & 2,252 & 4,013 & 214 & 3,306 & 48 \\
\quad Offload to 3x GPUs (2x 4090, 1x 3080 Ti)  & 9,624 & 373 & 205,957 & 2,278 & 3,825 & 191 & 3,241 & 49 \\
\bottomrule
\end{tabular}
\\[4pt]

\footnotesize {\textsuperscript{a}Denotes an edge-server configuration where window processing is offloaded from the Resource-constrained Edge
Devices. The server-side environment comprises one or more containerized nodes, each exclusively provisioned with an NVIDIA RTX 4090 or RTX 3080 Ti GPU, communicating with the client over a gigabit local area network.}
\end{table*}

%================ TABLE 2: GPU-BASED  ====================
\begin{table*}[!t]
\centering
% \small
\caption{Performance comparison across different hardware platforms when edge devices are configured with \textbf{GPU-enabled containers}. We report the average latency for the image encoder and the mask predictor. All times are measured in milliseconds (ms). 
\textbf{Abbreviations}: Enc. (Encoder), Pred. (Predictor).}
\label{tab:gpu_perf_comparison}
\setlength{\tabcolsep}{4.5pt}
\renewcommand{\arraystretch}{1.15}
\begin{tabular}{@{}l
            r@{\hspace{0.8em}}r
            r@{\hspace{0.8em}}r
            r@{\hspace{0.8em}}r
            r@{\hspace{0.8em}}r@{}}
\toprule
\multirow{2}{*}{\makecell[l]{\textbf{SAM Configuration} \\ \textbf{(GPU-based)}}} &
\multicolumn{2}{c}{\textbf{{\makecell{Jetson Orin\\Nano 8GB}}}} &
\multicolumn{2}{c}{\textbf{{\makecell{Jetson AGX Orin\\(Jetpack 5)}}}} &
\multicolumn{2}{c}{\textbf{{\makecell{Jetson AGX Orin\\ (Jetpack 6)}}}} &
\multicolumn{2}{c}{\textbf{RTX 3080 Ti}} \\
\cmidrule(lr){2-3}\cmidrule(lr){4-5}\cmidrule(lr){6-7}\cmidrule(lr){8-9}
& \textbf{Enc.} & \textbf{Pred.} & \textbf{Enc.} & \textbf{Pred.} & \textbf{Enc.} & \textbf{Pred.} & \textbf{Enc.} & \textbf{Pred.} \\
\midrule
Original SAM (Baseline)              & 2,704 & 256 & 1,577 & 42 & 1,034 & 191 & 256 & 138 \\
\midrule
\textit{Privacy-Enhanced SAM (Local Execution)}    & 2,759 & 260 & 1,569 & 34 & 1,059 & 191 & 259 & 139 \\
\midrule
\multicolumn{9}{@{}l}{\textit{Privacy-Enhanced SAM (Distributed Execution with Remote GPUs)\textsuperscript{a}}} \\
\quad Offload to 1x GPU (1x RTX 4090)           & 2,451 & 276 & 1,949 & 37 & 1,204 & 188 & 628 & 141 \\
\quad Offload to 2x GPUs (2x RTX 4090)          & 2,396 & 266 & 1,896 & 40 & 1,180 & 200 & 572 & 142 \\
\quad Offload to 3x GPUs (2x 4090, 1x 3080 Ti)  & 2,358 & 273 & 1,888 & 36 & 1,157 & 189 & 467 & 138 \\
\bottomrule
\end{tabular}
\\[5pt]
\footnotesize{\textsuperscript{a}Denotes an edge-server configuration where window processing is offloaded from the Resource-constrained Edge
Devices. The server-side environment comprises one or more containerized nodes, each exclusively provisioned with an NVIDIA RTX 4090 or RTX 3080 Ti GPU, communicating with the client over a gigabit local area network.}
\end{table*}

\vspace{0.3em}
\noindent\textbf{Evaluation Configurations.}
We benchmark performance across six configurations designed to isolate the impact of each architectural component and offloading strategy. The configurations are labeled as following in Tables~\ref{tab:cpu_perf_comparison} and~\ref{tab:gpu_perf_comparison}:

\begin{enumerate}[label=\textbf{(\arabic*)},]
  \item \textbf{Original SAM (CPU)}: Unmodified SAM in a CPU-only container.  
  \item \textbf{Original SAM (GPU)}: Unmodified SAM in a GPU-enabled container.  
  \item \textbf{PED-SAM (CPU)}: Two-staged ViT-B encoder on CPU.  
  \item \textbf{PED-SAM (GPU)}: Two-staged ViT-B encoder on GPU.  
  \item \textbf{PED-SAM (CPU → $k$ GPUs)}: CPU client offloads window encoding to $k$ remote RTX 4090 servers.  
  \item \textbf{PED-SAM (GPU → $k$ GPUs)}: GPU client offloads window encoding to $k$ remote RTX 4090 servers.  
\end{enumerate}

The distributed configurations, (5) and (6), are architected for horizontal scalability. Consequently, the total encoder latency, which is bound by the parallel processing of the window batch, should continue to improve as the pool of available External Servers expands.  

Our experiment setup was constrained by the number of high-end GPUs we have access to. As a result, configurations (5) and (6) were limited to offloading computation to at most 3 GPUs acting as external servers. However, our results demonstrate that the design is horizontally scalable, and its performance would further improve with access to a larger number of resources.

\begin{table*}[t]
\centering
\caption{Segmentation mIoU on COCO val2017 for Privacy-Enhanced SAM (PED-SAM) Compared to Original SAM and Input-Level Privacy Method
}
\label{tbl:miou_fullcomp}
% \small
\setlength{\tabcolsep}{8pt} 
\renewcommand{\arraystretch}{1.2} 
\scalebox{0.98}{
\begin{tabular}{c|ccc}
\toprule
\textbf{Models} & ViT-Huge & ViT-Large & ViT-Base \\
\midrule
Original SAM models (unmodified inputs)   & 0.584 & 0.582 & 0.534 \\
Original SAM (input processed with basic obfuscation) & 0.209 & 0.193 & 0.189 \\
\textbf{Our Privacy-Enhanced SAM (unmodified inputs, original model weights)}  &  0.563 & 0.553 & 0.523\\
\bottomrule
\end{tabular}}
% \vspace{-0.5em}
\end{table*}

\vspace{0.3em}

\noindent\textbf{Latency Performance Results.}
Tables~\ref{tab:cpu_perf_comparison} and~\ref{tab:gpu_perf_comparison} present the end-to-end encoder and predictor latencies that establish our performance baselines. Our analysis of the original SAM model reveals a critical characteristic of its lightweight mask predictor: unlike the heavily parallelized encoder, its latency is governed by serial, single-thread processing speed rather than massively parallel throughput. This is most evident on the Ryzen 9 desktop, where the CPU's high-frequency cores complete the predictor task in just 46 ms,  outperforming the powerful RTX 3080 Ti GPU (138 ms).

In contrast, on Jetson-class SoCs the performance characteristics differ.
Under JetPack~5, the AGX~Orin’s CPU execution was limited by a one-core scheduling constraint, leading to significantly inflated encoder latency (over 400{,}000\,ms in Table~\ref{tab:cpu_perf_comparison}), which we include for completeness.
With JetPack~6 enabling all 12 cores, the AGX Orin’s CPU achieves 7{,}899\,ms for the encoder and 188\,ms for the predictor, compared to 1{,}034\,ms and 191\,ms on its integrated GPU. Thus, while the GPU retains a clear advantage for the encoder ($\sim$7.6$\times$ faster), the predictor now runs equally well on the CPU or GPU. On the Orin Nano, the GPU is consistently faster for both stages (2{,}704/256\,ms vs.\ 17{,}698/444\,ms). These results highlight a key optimization opportunity that our framework can exploit: an intelligent client can always schedule the encoder on the fastest accelerator while dynamically selecting the predictor’s placement (CPU or GPU) to minimize end-to-end latency. When remote GPUs are available, distributed execution further reduces encoder latency while leaving predictor performance unchanged.

Our analysis also confirms that restructuring the ViT encoder into our two-stage, privacy-enhancing design imposes negligible additional cost. Across all local test cases on both CPU and GPU, the added latency remains below 6\% on the encoder. For example, in CPU-only mode the encoder overhead is +0.6\% on the Ryzen~9, +0.01\% on the Orin Nano, and +4.3\% on the AGX Orin (JetPack~6). On GPU-based execution, the overhead is similarly minor (e.g., Orin AGX: 1{,}034\,ms$\rightarrow$1{,}059\,ms; RTX~3080\,Ti: 256\,ms$\rightarrow$259\,ms). For the predictor, overhead is negligible on GPUs (e.g., AGX Orin: 191\,ms$\rightarrow$191\,ms), but can be higher if the predictor is run on the CPU (188\,ms$\rightarrow$254\,ms). Because our framework allows the client to flexibly assign the predictor to the faster resource per platform, the overall privacy-enhancing restructuring does not introduce a significant performance bottleneck.

Our framework’s primary benefit is realized in distributed configurations, where offloading the encoder stage to external GPUs yields substantial performance gains for resource-constrained clients. In CPU-only mode, the improvements are most significant on Jetson-class SoCs: encoder latency is reduced by up to 51.5\% on the AGX Orin (7{,}899\,ms $\rightarrow$ 3{,}825\,ms) and 44.0\% on the Orin Nano (17{,}698\,ms $\rightarrow$ 9{,}624\,ms). Even the powerful desktop Ryzen 9 CPU benefits, dropping from 4{,}135\,ms to 3{,}241\,ms ($\sim$21.6\%). 
When offloading from GPU-enabled edge devices, the trade-offs become clearer. 
The Orin Nano's modest integrated GPU sees meaningful reductions (2{,}704\,ms $\rightarrow$ 2{,}358\,ms, $\sim$12.8\%), but for the more powerful AGX Orin GPU (1{,}034\,ms baseline) and RTX~3080\,Ti (256\,ms baseline), the communication and serialization costs outweigh the savings from additional parallelism, leading to
similar or slightly higher latency.
This underscores a predictable system trade-off: offloading is most effective when the remote compute power vastly exceeds the client’s local capabilities, and the network overhead remains low.

In summary, our empirical evaluation demonstrates that PED-SAM successfully achieves its design goals. It enhances privacy with minimal computational overhead, and by strategically harnessing remote resources, it dramatically accelerates the heavyweight encoder stage. The results confirm that the framework delivers the most profound benefits to computationally constrained edge devices, making large-scale ViT models like SAM practical in latency-sensitive and resource-limited scenarios.

\vspace{0.3em}
\noindent\textbf{Vision Task Performance Results.}
We further present the segmentation performance in Table \ref{tbl:miou_fullcomp}. 
These mean Intersection over Union (mIoU) results were obtained on the COCO val2017 dataset~\cite{lin2014microsoft} using the evaluation methodology from~\cite{segment-anything-fast}. 
In this methodology, for each ground truth mask, its ``center point'' was used as a point prompt, and the model's predicted mask was then compared against the original ground truth. 

Our results show that the privacy-enhanced SAM model achieved mIoU scores comparable to the original SAM model across various SAM ViT configurations. The privacy-enhanced SAM models achieved mIoU scores of 0.56, 0.55, and 0.52 for ViT-Huge, ViT-Large, and ViT-Base, slightly lower than the original SAM scores of 0.58, 0.58, and 0.53, respectively. 
This close alignment in performance indicates that adapting the two-staged design to enable privacy-enhanced distributed processing does not significantly impact image segmentation accuracy. 
We also want to point out that when we adopted the two-staged design, we used the layer weights directly without any changes, even if the layer has been changed  from a global layer to a local layer or
vice versa. 
Based on the results reported in ViTDet~\cite{li2022exploring}, 
we anticipate the performance to be improved if the ViT model is finetuned. 

\vspace{1mm}
%{\color{blue}
\noindent\textbf{Comparison with Input-Level Baseline.}
To establish a baseline for input-level privacy defenses, we evaluated Gaussian obfuscation (blur) applied directly to input images. We preprocessed COCO images with this blur and then ran the unchanged SAM variants (no retraining) on the processed inputs. The segmentation and mIoU computation pipeline remained identical across all experimental conditions. 

As shown in Table~\ref{tbl:miou_fullcomp}, Gaussian obfuscation substantially degrades segmentation quality, whereas our approach maintains high utility.
Basic Gaussian blur reduces ViT-H mIoU from 0.584 to 0.209 because it removes the high-frequency spatial cues that SAM relies on for precise mask delineation.
By contrast, our partition-based approach preserves near-baseline utility (ViT-H mIoU: 0.584 $\rightarrow$ 0.563) without changing the input representation or retraining the model. 
By constraining each server's view to isolated partitions while maintaining full pixel fidelity, our method offers a practical, deployment-ready solution that balances privacy protection with task utility for resource-constrained edge devices.
%}%blue

\subsection{Privacy Evaluation}
\noindent\textbf{Dataset and Image Selection.}
For privacy-focused experiments, we selected 12 diverse images from the COCO dataset, encompassing both indoor and outdoor environments with sub-categories including people, animals, and objects. This selection ensures comprehensive coverage of real-world scenarios, enabling us to evaluate the framework's robustness across various contexts.

\vspace{0.3em}
\noindent\textbf{Partition Settings and Reconstruction Methods.}
To evaluate privacy protection, images were partitioned into tiles of varying sizes using three different partition schemes: $2\times2$, $4\times 3$, and $5\times 5$. 
We assume an adversary at the cloud server can use the following two state-of-the-art methods for image reconstruction:

\begin{itemize}
    \item \textbf{ViTMAE (Vision Transformer Masked Autoencoder)~\cite{he2022masked}}: A self-supervised learning technique that reconstructs images from latent representations by predicting masked portions based on visible patches.
    \item \textbf{Adobe Firefly (Diffusion Network)~\cite{adobefirefly}}: An advanced AI tool from Adobe, capable of prompt-less inpainting. It understands image context and style to seamlessly fill missing areas or extend image boundaries without explicit text prompts.
\end{itemize}

The ViTMAE architecture supports only the $14\times 14$ patching scheme, i.e., a total of 196 patches. For each $m\times n$ partition scheme, we set each partition to contain $\lfloor 14/m \rfloor \times \lfloor 14/n \rfloor$ patches. 
During experiments, for each partition, we mask the remaining patches and use ViTMAE to predict the pixel values of these masked patches.
For a direct comparison, we apply Adobe Firefly inpainting to the same partitions as ViTMAE, i.e., identical partition scheme and window extents.

\begin{figure}[!t]
    \centering
    \begin{subfigure}[t]{0.48\textwidth}
        \centering
\includegraphics[width=0.5\linewidth]{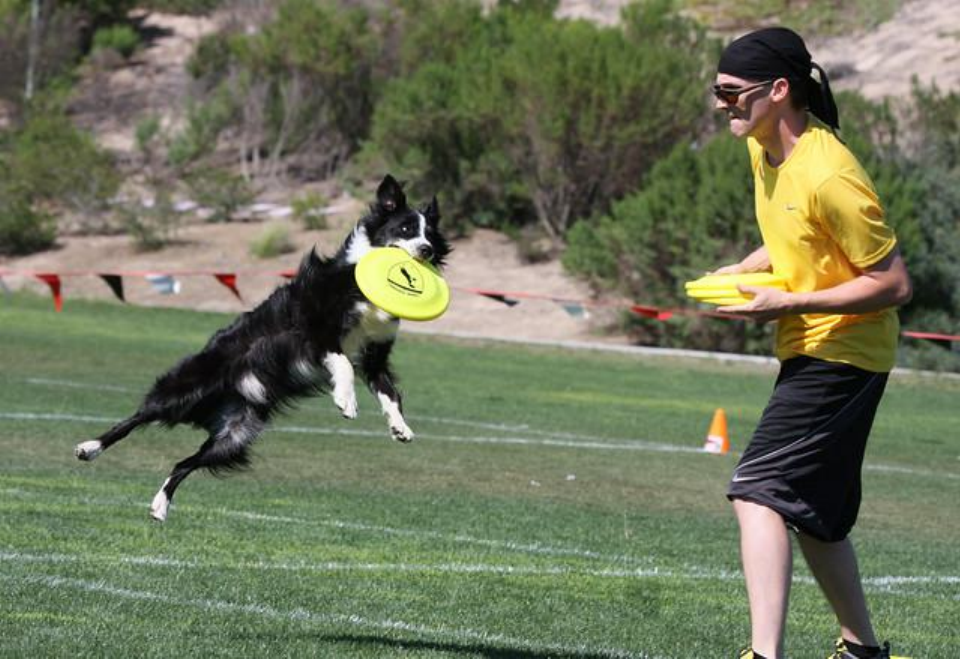}
        \caption{Original Image}
    % \label{fig:og}
    \end{subfigure}    
    \vspace{0.3cm}

    \begin{subfigure}[t]{0.48\textwidth}
        \centering        \includegraphics[width=\linewidth]{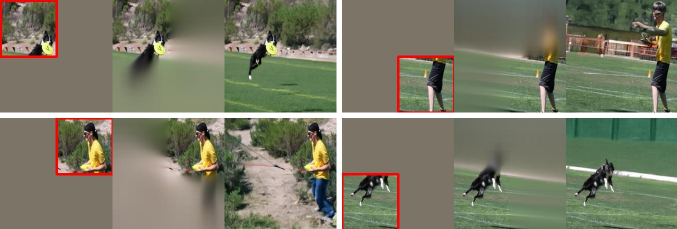}
        \caption{$2\times 2$ partitions}
        % \label{fig:2x2seg}
    \end{subfigure}
    \vspace{0.3cm}

    \begin{subfigure}[t]{0.48\textwidth}
        \centering
        \includegraphics[width=\linewidth]{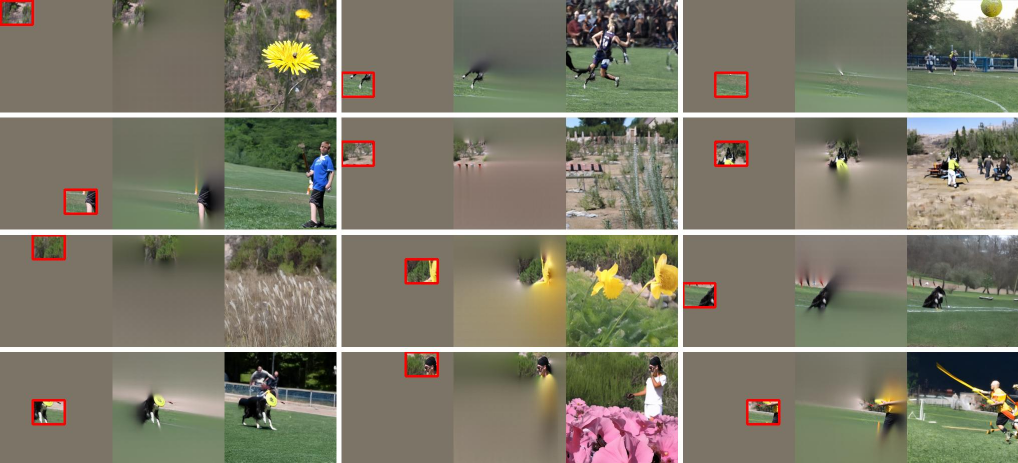}
        \caption{$4\times 3$ partitions}
        % \label{fig:4x3seg}
    \end{subfigure}

    \caption{\textmd{Visualization results of reconstructed image from \textbf{ViTMAE} \cite{he2022masked} and \textbf{Adobe Firefly} \cite{adobefirefly} using the $2\times 2$ and $4\times 3$ partition schemes. Figure \ref{fig:visual_comp}(a) shows the original image. In each triplet in Figure \ref{fig:visual_comp}(b) and Figure \ref{fig:visual_comp}(c), we present the masked image with highlighted window partition (left), i.e.,  visual data shared with the cloud server, ViTMAE reconstruction (middle), and Adobe Firefly reconstruction (right). }
    }
    \label{fig:visual_comp}
\end{figure}

\begin{figure*}[!t]
    \centering
    \includegraphics[width=\linewidth]{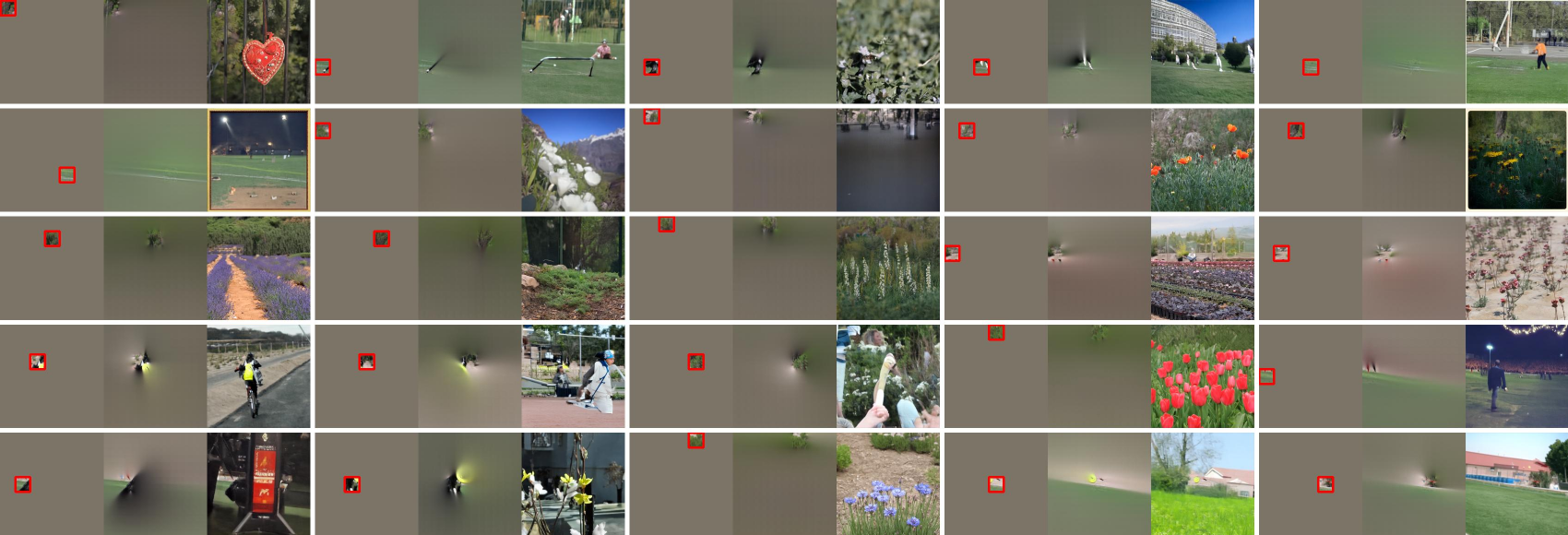}
    \caption{\textmd{Visualization results of reconstructed images from \textbf{ViTMAE} \cite{he2022masked} and \textbf{Adobe Firefly} \cite{adobefirefly} using the $5\times 5$ partition scheme. For each triplet, we show the masked image where the window partition is highlighted (left), ViTMAE reconstruction (middle), and Adobe Firefly reconstruction (right). The original image is shown in Figure \ref{fig:visual_comp}(a). We can observe that given a small window partition (high protection ratio), the state-of-the-art image reconstruction methods cannot restore the original images, indicating our privacy-enhanced ViT can protect the image privacy effectively.}
    }
    \vspace{-0.5em}
    \label{fig:5x5_partition}
\end{figure*}

Sample visual results can be found in Figures \ref{fig:visual_comp} and \ref{fig:5x5_partition}.
With the $2\times 2$ partition scheme, each partition retains much information in the original image, e.g., the person and the dog. The images reconstructed by ViTMAE and Adobe Firefly also contain such information.
With the $5 \times 5$ partition scheme, however, as partition size decreases, the reconstructed images increasingly diverge in information from the original image. 

\vspace{0.3em}
\noindent\textbf{Image Reconstruction Quality Metrics.}
We evaluated the quality of image reconstructions with different partition settings using two pixel-level similarity metrics:

\begin{itemize}
    \item \textbf{Structural Similarity (SSIM)~\cite{wang2004image}}: This metric assesses the similarity between the original and reconstructed images. A lower SSIM score indicates less similarity, suggesting better privacy protection, as the reconstructed image deviates more from the original.
    \item \textbf{Mean Absolute Error (MAE)}: This metric measures the average absolute pixel-level difference between the original and reconstructed images. A higher MAE score means greater differences, enhancing privacy protection by making reconstruction less accurate.
\end{itemize}

Tables \ref{tab:ssim_mae_metrics} and \ref{tab:adob_ssim_mae_metrics} show that for both reconstruction methods, increasing the number of partitions reduces the similarity between the reconstructed and original images (lower SSIM) while increasing pixel-level differences (higher MAE). 
Notably, the $5\times 5$ partition setting achieves the lowest \textbf{SSIM} and highest \textbf{MAE} scores for both reconstruction methods, indicating the least similarity to the original images. This suggests that finer partitioning effectively obfuscates image content, thus enhancing privacy protection.

%%%%% protection ratio %%%%%%
\vspace{0.3em}
\noindent\textbf{Worst-Case Scenario Analysis.}
In the event of a data breach, our partitioning approach limits exposure to no more than $1/25$ of the image  when using the $5 \times 5$ partition scheme. This \textit{protection ratio} ensures that individual image pieces are insufficient for meaningful reconstruction of the original image, thus effectively protects user privacy.

\begin{table}[!t]
\centering
    \caption{SSIM and MAE Results for ViTMAE~\cite{he2022masked} Reconstructions}
    \label{tab:ssim_mae_metrics}
    \begin{tabular}{c|c|c}
        \toprule
        \textbf{Partition Scheme} & \textbf{Avg. SSIM$\downarrow$} & \textbf{Avg. MAE$\uparrow$} \\
        \midrule
        2\(\times\)2 & 0.512 & 0.130 \\
        4\(\times\)3 & 0.376 & 0.172 \\
        \textbf{5\(\times\)5} & \textbf{0.336} & \textbf{0.192} \\
        \bottomrule
    \end{tabular}
\end{table}

\begin{table}[!t]
    \centering
    \caption{SSIM and MAE Results for Adobe Firefly~\cite{adobefirefly} Reconstructions}
    \label{tab:adob_ssim_mae_metrics}
    \begin{tabular}{c|c|c}
        \toprule
        \textbf{Partition Scheme} & \textbf{Avg. SSIM$\downarrow$} & \textbf{Avg. MAE$\uparrow$} \\
        \midrule
         2\(\times\)2 & 0.420 & 0.166 \\
        4\(\times\)3 & 0.257 & 0.230 \\
        \textbf{5\(\times\)5} & \textbf{0.206} & \textbf{0.260} \\ 
        \bottomrule
    \end{tabular}
    \vspace{-1em}
\end{table}

%%%%% yolo result %%%%%%
\vspace{0.3em}
\noindent\textbf{Downstream Task Evaluation.}
To further evaluate the effectiveness of our privacy protection mechanism, we evaluated the quality of reconstructed images using YOLOv9~\cite{wang2024yolov9}, a state-of-the-art object detection model known for its accuracy and efficiency. Specifically, YOLOv9-C was selected for its performance on the MS COCO dataset, achieving an Average Precision (AP$_{val}$) of 53.0\%, AP$_{50val}$ of 70.2\%, and AP$_{75val}$ of 57.8\% with 25.3 million parameters and 102.1 billion FLOPs. This robust performance makes YOLOv9-C an ideal benchmark for assessing the reliability of reconstructed images in downstream tasks.
In our evaluation, we focused on four object-level privacy metrics to measure object detection quality on reconstructed images:

\begin{itemize}
    \item \textbf{Precision:} The proportion of correctly identified objects out of all detected objects.
    \item \textbf{Recall:} The proportion of actual objects that were successfully detected.
    \item \textbf{Intersection over Union (IoU):} Measures the overlap between the predicted bounding box and the ground truth.
    \item \textbf{Average Matched Detections:} The average number of detections that successfully match the ground truth.
\end{itemize}

To ensure reliability, we filtered out detections with confidence scores below 75\%. This threshold eliminated low-confidence detections that could affect the analysis, allowing only high-confidence results to be analyzed. Precision and Recall were used as the primary metrics for evaluating YOLOv9-C's performance on reconstructed images with different partition settings.
The results are shown in Tables \ref{tab:vitmae_summary} and \ref{tab:adobegen_summary}.

For the ViTMAE reconstruction method, the $2\times 2$ partitioning resulted in a precision of 0.167 and a recall of 0.143, indicating a moderate similarity to the original images with some detectable objects. As the partition grid was increased to $4 \times 3$, precision and recall further declined to 0.074 and 0.034, respectively, reflecting a significant decrease in detection performance and enhanced privacy protection. At $5\times 5$, precision and recall dropped to 0.000, marking a complete breakdown in object detection capabilities. This substantial decline demonstrates that $5\times 5$ partitioning effectively obfuscates image content, ensuring robust privacy protection.

\begin{table}[!t]
    \centering
    \small
    \caption{YOLOv9 Detection Result for Images Restored by ViTMAE Across Different Partition Schemes}
    \label{tab:vitmae_summary}
    \begin{tabular}{cccccc}
        \toprule
        \makecell{\textbf{Partition}\\\textbf{Scheme}} &
        \makecell{\textbf{Precision$\downarrow$}} &
        \makecell{\textbf{Recall$\downarrow$}} &
        \makecell{\textbf{IoU$\downarrow$}} &
        \makecell{\textbf{Avg. Matched}\\\textbf{Detections $\downarrow$}} \\
        \midrule
        2\(\times\)2 & 0.167 & 0.143 &  0.165 & 0.396 \\
        4\(\times\)3 & 0.074 & 0.034 &  0.044 & 0.083 \\
        \textbf{5\(\times\)5} & \textbf{0.000} & \textbf{0.000} &  \textbf{0.000} & \textbf{0.000} \\
        \bottomrule
    \end{tabular}
    % \vspace{-1em}
\end{table}

\begin{table}[!t]
    \centering
    \small
    \caption{YOLOv9 Detection Result for Images Restored by Adobe Firefly Across Different Partition Schemes}
    \label{tab:adobegen_summary}
    \begin{tabular}{cccccc}
        \toprule
        \makecell{\textbf{Partition}\\\textbf{Scheme}}  &
        \makecell{\textbf{Precision$\downarrow$}} &
        \makecell{\textbf{Recall$\downarrow$}} &
        \makecell{\textbf{IoU$\downarrow$}} &
        \makecell{\textbf{Avg. Matched}\\\textbf{Detections $\downarrow$}} \\
        \midrule
        2\(\times\)2 & 0.207 & 0.178 & 0.211 & 0.448 \\
        4\(\times\)3 & 0.044 & 0.013 & 0.025 & 0.052 \\
        \textbf{5\(\times\)5} & \textbf{0.007} & \textbf{0.006} & \textbf{0.006} & \textbf{0.010}\\
        \bottomrule
    \end{tabular}
    \vspace{-0.5em}
\end{table}
Similarly, for the Adobe Firefly reconstruction method, $2\times 2$ partitioning achieved a precision of 0.207 and recall of 0.178, suggesting that reconstructed images retained some detectable features, albeit at a reduced fidelity compared to the originals. 
Increasing the number of partitions to $4\times 3$ further decreased precision and recall to 0.044 and 0.013, respectively, indicating enhanced privacy protection through diminished detection performance. 
The $5\times 5$ partition scheme resulted in minimal precision (0.007) and recall (0.006), effectively nullifying the model's ability to detect objects accurately. 
These negligible values underscore the efficacy of the $5\times 5$ partitioning in preventing reliable object detection from reconstructed images.

\begin{figure}[!t]
\centering
% Second 2x2 figure
\begin{subfigure}[t]{0.49\columnwidth}
    \includegraphics[width=\linewidth]{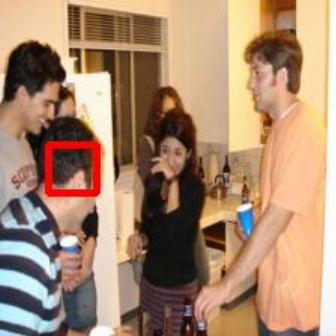}
    \caption{Original image with a partition in the $5\times 5$ partition scheme highlighted. }
    \label{fig:5}
\end{subfigure}\hfill
\begin{subfigure}[t]{0.49\columnwidth}
    \includegraphics[width=\linewidth]{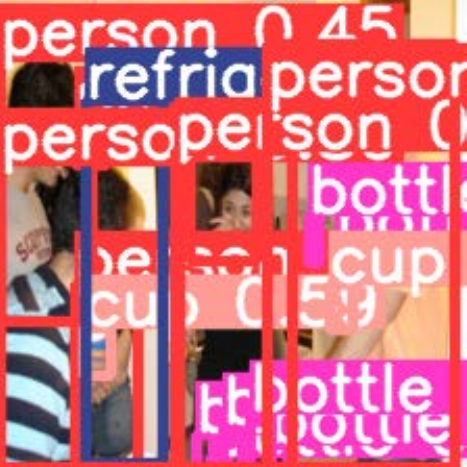}
    \caption{YOLO detections on the original image.}
    \label{fig:6}
\end{subfigure}

\medskip
\begin{subfigure}[t]{0.49\columnwidth}
    \includegraphics[width=\linewidth]{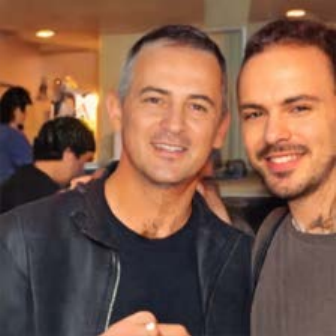}
    \caption{Adobe Firefly reconstruction from a single partition.}
    \label{fig:7}
\end{subfigure}\hfill
\begin{subfigure}[t]{0.49\columnwidth}
    \includegraphics[width=\linewidth]{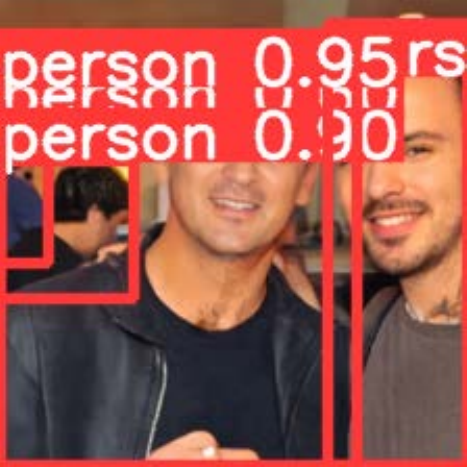}
    \caption{YOLO detections on the reconstructed image, which differs significantly from the original.}
    \label{fig:8}
\end{subfigure}
\vspace{-0.5em}
\caption{\textmd{Visualization at high recall and IoU using a partition from the $5\times 5$ partition scheme. Although YOLO detects the same object class (e.g., person) in both original and reconstructed images, the significant differences in the detected information demonstrate effective privacy protection.}}
\vspace{-1em}
\label{fig:second}
\end{figure}

An additional observation from the YOLO evaluation revealed instances where reconstructed images exhibited higher IoU and recall scores despite the apparent degradation in image quality, as indicated by low SSIM and high MAE values. This anomaly arises not from successful content reconstruction but rather from the presence of different or unrelated content within the reconstructed images that YOLO mistakenly classifies as existing object classes. For example, larger or multiple bounding boxes of the same class in reconstructed images may inadvertently yield higher IoU and recall scores. However, upon visual inspection, these detections do not correspond to the original image content, as evidenced by low SSIM (e.g., 0.1704) and high MAE (e.g., 0.3032) values in Figure \ref{fig:second}. 
This discrepancy further validates that reconstructed images do not retain meaningful or accurate content, which shows the robustness of our privacy protection mechanism.

\section{Limitations \& Future Work}

\noindent\textbf{Adaptive partitioning for sensitive regions.}
Our current design uses a fixed, uniform windowing policy, which already reduces reconstruction and detection risks as the number of partitions increases.
We acknowledge a residual worst case that can occur
when a small sensitive object (e.g., a full face, or credit card number) falls entirely inside a single window. 
A potential direction for future work is to explore content-aware, adaptive partitioning.  
% a lightweight image content sensitivity estimator on the edge device (e.g., a compact face/text detector or a lightweight PII classifier \cite{octopii, octopii_blog},
% or a lightweight classifier trained on face and text) to drive \emph{adaptive partitioning}. 
% That is, flagged regions are subdivided into finer windows, while non-flagged regions remain the default window size. 
Such a policy introduces predictable trade-offs, such as added local computation and  increased communication overhead. 
Future evaluation would be needed to determine how to best reassemble heterogeneous outputs prior to global attention layers to maintain downstream task performance.

\vspace{2mm}
\noindent\textbf{Video frame assembly risk.}
Extending image partitioning to video frames introduces a temporal attack surface: an adversary server might aggregate tiles observed across frames and partially ``stitch'' content that is unrecoverable from any single frame. 
This suggests that a simple frame-by-frame partitioning scheme is insufficient for videos. Future research will need to investigate temporal-aware partitioning schemes to mitigate this cross-frame assembly risk.

\vspace{2mm}
\noindent\textbf{Utility recovery.}
Finally, we expect that the small utility drop observed relative to unpartitioned baselines (Table \ref{tbl:miou_fullcomp}) can be recovered by fine-tuning the ViT backbone on the partitioned data. 
Since our current evaluation used the original model weights without any re-training, future work could explore this fine-tuning to potentially recover the minor performance drop.

\section{Conclusion}
In this paper, we introduced a  privacy-enhanced distributed edge-cloud offloading framework for Vision Transformers that prioritizes user privacy protection. By partitioning images into smaller tiles and processing them across external servers, the framework ensures that adversaries cannot reconstruct enough meaningful content even in the event of a data breach. 
This framework offers a practical solution for scalable ViT-based vision processing while substantially reducing private content leakage by storing and processing the full image only on the local device and only transmitting a partition of the image
for external processing. This design is well-suited for integration into modern smartphone and wearable AR/VR/MR devices, providing an additional layer of user content protection.

\begin{acks}
We appreciate constructive comments from anonymous referees. 
The kitchen source image used in Figures \ref{fig:poc-results}, \ref{fig:global_local}, and \ref{fig:poc} is courtesy of Jason Briscoe, via Unsplash.
This work is partially supported by NSF under grants 2200048 and 2350188. 
\end{acks}

\balance
\bibliographystyle{ACM-Reference-Format}
\bibliography{privacy}
\end{document}